\documentclass[manuscript]{aastex}

\usepackage{graphicx}
\usepackage{epsfig}
\usepackage{times}
\usepackage{natbib}
\usepackage{url}
\usepackage{color}
\usepackage{amssymb,amsmath}
\shorttitle{Filament eruptions and EIT wave during B-class
flare}
\shortauthors{}

\begin{document}

\title{Peculiar Stationary EUV Wave Fronts in the eruption on 2011 May 11}

\author{R. Chandra}
\affil{Department of Physics, DSB Campus, Kumaun University, Nainital -- 263 001, India}
\email{rchandra.ntl@gmail.com}
\author{P. F. Chen}
\affil{School of Astronomy \& Space Science, Nanjing University, Nanjing -- 210 023, China}
\author{A. Fulara}
\affil{Department of Physics, DSB Campus, Kumaun University, Nainital -- 263 001, India}
\author{A. K. Srivastava}
\affil{Department of Physics, Indian Institute of Technology (BHU), Varanasi -- 221 005, India}
\author{W. Uddin}
\affil{Aryabhatta Research Institute of Observational Sciences, Manora Peak, Nainital -- 263 002, India}

\begin{abstract}
\noindent {We present and interpret the observations of extreme ultraviolet (EUV) waves associated with a filament eruption on 2011 May 11.
The filament eruption also produces a small B-class two ribbon flare and a coronal mass ejection (CME). The event is observed by
the {\em Solar Dynamic Observatory} ({\it SDO}) with high spatio-temporal
resolution data recorded by Atmospheric Imaging Assembly (AIA). As the filament erupts, we observe two types of EUV
waves (slow and fast) propagating outwards. The faster EUV wave has a propagation velocity of $\sim$500 km s$^{-1}$ and
the slower EUV wave has an initial velocity of $\sim$120 km s$^{-1}$. We report for the
first time that not only the slower EUV wave stops at a magnetic separatrix to
form bright stationary fronts, but also the faster EUV wave transits a magnetic
separatrix, leaving another stationary EUV front behind.}

\end{abstract}

\section{Introduction}

As the two largest eruptive phenomena in solar atmosphere, both solar flares
and coronal mass ejections (CMEs) are frequently associated with erupting
filaments, which later become the core of the CMEs. These three phenomena are the
three important ingredients in the standard CSHKP model \citep{Carmichael64,Sturrock66, Hirayama74, Kopp76} for
flare/CME. In this unified model, the erupting filament plays a very
crucial role \citep{Chen11a}.

The erupting filament also drives wave phenomena in the solar atmosphere,
which are manifested in radio, H$\alpha$, extreme ultraviolet (EUV), and other
wavelengths. In contrast to H$\alpha$ Moreton waves which occur sparsely, the
frequently accompanied waves that can be directly imaged are {\it EIT waves}.
EIT waves were discovered in the EUV difference images with the EUV imaging
telescope \citep[EIT,][]{delab1995} on board the {\em Solar and Heliospheric
Observatory} \citep[\textit{SOHO},][]{Domingo95} by \citet{Moses97} and
\citet{Thompson98}. This is the reason why they were named EIT waves more than
17 years ago. Later, several other names were invented for this phenomenon,
such as EUV waves and large-scale coronal propagating fronts \citep{Nitta13}.
EIT waves were initially proposed to be the coronal counterparts
of H$\alpha$ Moreton waves, i.e., fast-mode magnetohydrodynamic (MHD) waves in
the solar corona \citep{Thompson98, Wang00, Wu01}. However, these waves
present some features which cannot be easily explained by the fast-mode wave
model. For example, according to \citet{Klassen00}, the typical velocity of
EIT waves is in the range of 170--350 km s$^{-1}$, which is about 3 or more
times slower than Moreton waves. In some cases, the EIT wave speed can be as
small as $\sim$10 km s$^{-1}$ \citep{zhuk09}. In order to resolve the velocity
discrepancy, several non-wave models have also been proposed \citep{Gallagher11, chen12, pats12, Liu14}.
For more reviews on EUV waves, see \citet{warmuth07}, \citet{wills09}, \citet{zhukov07}, \citet{zhukov11}, \citet{warmuth11}, and \citet{patsourakos12}. Very recently 
\citet{warmuth15} presented a very excellent review on the globally propagating coronal waves. The review is
focussed on various observational findings, physical nature and different models of EUV waves proposed in the past years. It seems now that there should be two
types of EUV waves with different velocities, and the EIT wave initially
discovered by \citet{Moses97} and \citet{Thompson98} may correspond to the
slower type of EUV waves \citep{Chen02}. Following \citet{chen12}, we use
``EIT waves'' for the slower type of EUV waves specifically. \citet{biesecker02} did a statistical study of EIT waves 
observed by {\em SOHO}/EIT (195 \AA) telescope and found that some of the EIT waves have sharp bright features, which they called ``S-waves''. 
These S-waves may be the signature of Moreton waves. They also concluded that 
EIT waves having S-shape signatures are always associated with both flares and 
CMEs. On the basis of {\em SOHO}/EIT observations, \citet{Zhukov04} suggested 
a bimodal characteristic for EIT waves, i.e. including a wave mode and an 
eruptive mode component. The wave-mode component is a wavelike phenomenon and
represented by pure MHD waves. The eruptive-mode component is defined as 
propagating bright fronts and dimming as a result of 
successive stretching of field lines during the eruption of CMEs as modeled by \citet{Chen02}. \citet{Downs12} presented a comprehensive observations of 
2010 June 13 EUV wave observed by {\it SDO}/AIA in different channels and conducted a 3D MHD simulations of CME eruptions and the associated EUV waves. 
They suggested that the outer component of EUV waves behaves as a fast-mode wave and found that this component later decouples 
from the associated CME. Their study distinguishes between wave and and non-wave mechanisms of EUV waves.

An even more serious issue that led
\citet{Delannee99} and \citet{Delannee00} to doubt the fast-mode wave
model for the EIT waves is that they found stationary wave fronts in several
events. These stationary fronts are found to be located at magnetic separatrix,. Considering that this
feature can hardly be accounted for by the fast-mode wave model, they related
the stationary EUV front to the opening of the closed magnetic field lines
during the CME. The stationary EUV front was also explained in the framework of
the magnetic field line stretching model proposed by \citet{Chen02}. With
numerical simulations, \citet{Chen05} and \citet{chen06a} illustrated how a
propagating EIT wave stops at the magnetic separatrix. However, as mentioned
by \citet{Delannee99}, although it is unlikely, there is a possibility that
the stationary EUV front is an artifact because it might be due to successive
wave fronts reaching the same location after $\sim$15 min, which is the
cadence of the EIT observations. With the high cadence of the {\em SDO}/AIA
observation up to 12 s, this issue can be settled conclusively.

With the purpose to better understand the EUV wave and its stationary
fronts, in this paper we present our study of the filament eruption event
originating between the active regions NOAA 11207  and 11205 on 2011 May 11.
The paper is organized as follows: Section \ref{obs} describes the
instruments and the observational data. The observational view of filament 
eruption and associated phenomena are investigated in Section \ref{filament}, 
whereas the EUV waves and its stationary fronts are analyzed in Section 
\ref{euv}. The discussion of our results is presented in Section 
\ref{discussion}. Finally, the conclusion is drawn in Section \ref{sum}.

\section{Instrumentation and Data}
\label{obs}

The Atmospheric Imaging Assembly \citep[AIA,][]{Lemen12} on board the
{\it SDO} satellite \citep{Pesnell12} observes the full Sun with different
filters in EUV and UV spectral lines with a cadence up to 12 s and a pixel
size of $0\farcs 6$. For this current study, we use the AIA 171 \AA, 193
\AA, and 304 \AA\ data. The high cadence and high spatial resolution of the
AIA images allow us to see more details of the filament eruption and the
associated EUV waves. To have a better view of EUV waves, we utilize the base
difference images by subtracting each image with one before eruption. All the
images are corrected for the solar differential rotation. For the
magnetograms, we use the data observed by the Helioseismic and Magnetic Imager
 \citep[HMI,][]{sche12} aboard {\it SDO}. HMI measures the photospheric
magnetic field of the Sun with a cadence of 45 sec and a spatial resolution
1\arcsec.

\begin{figure}[ht]
\centering
  \begin{minipage}{\columnwidth}
  \centering
\hbox
{
\includegraphics[width=5.6cm]{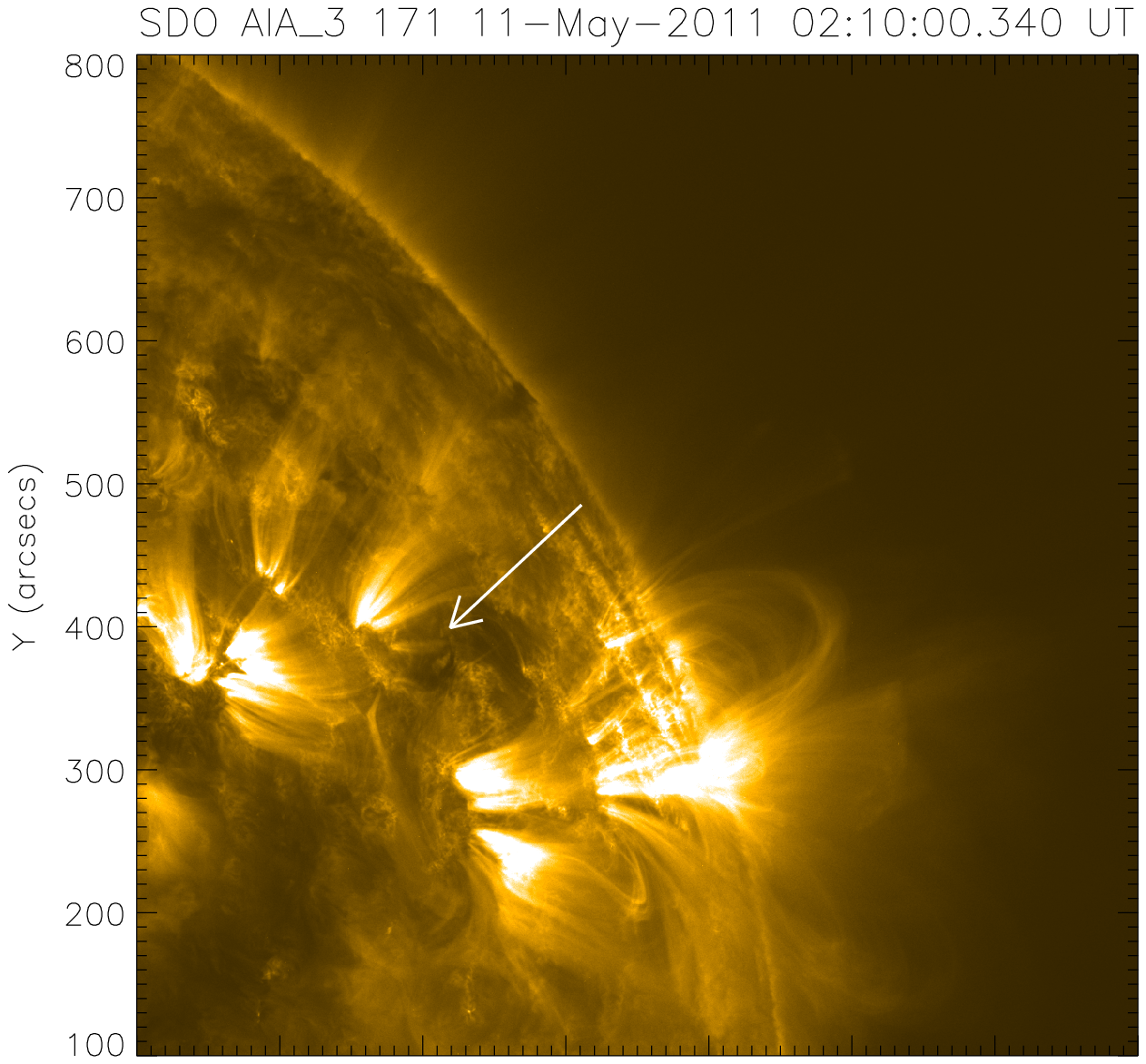}
\hspace{2mm}
\includegraphics[width=5cm]{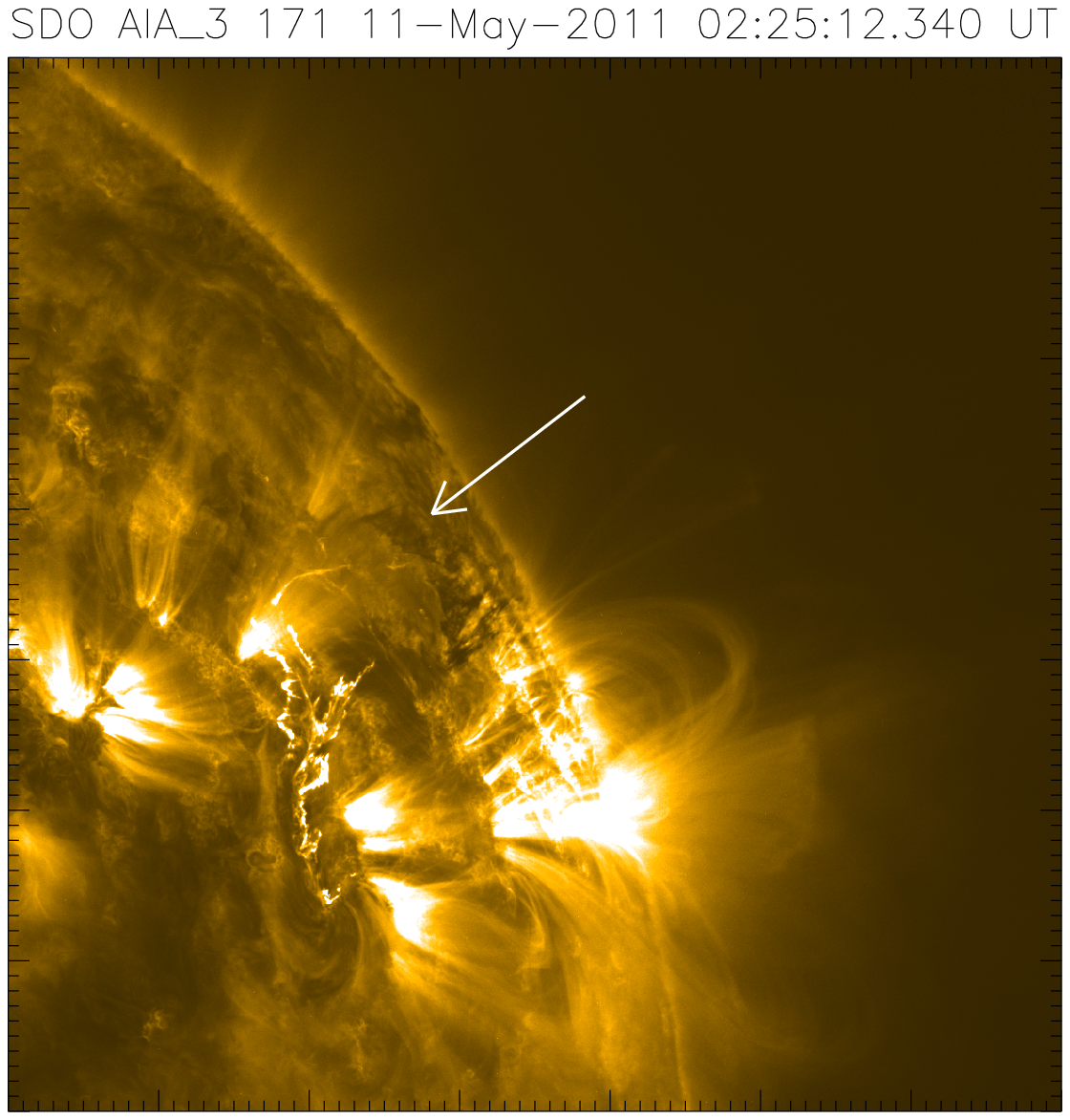}
\hspace{2mm}
\includegraphics[width=5cm]{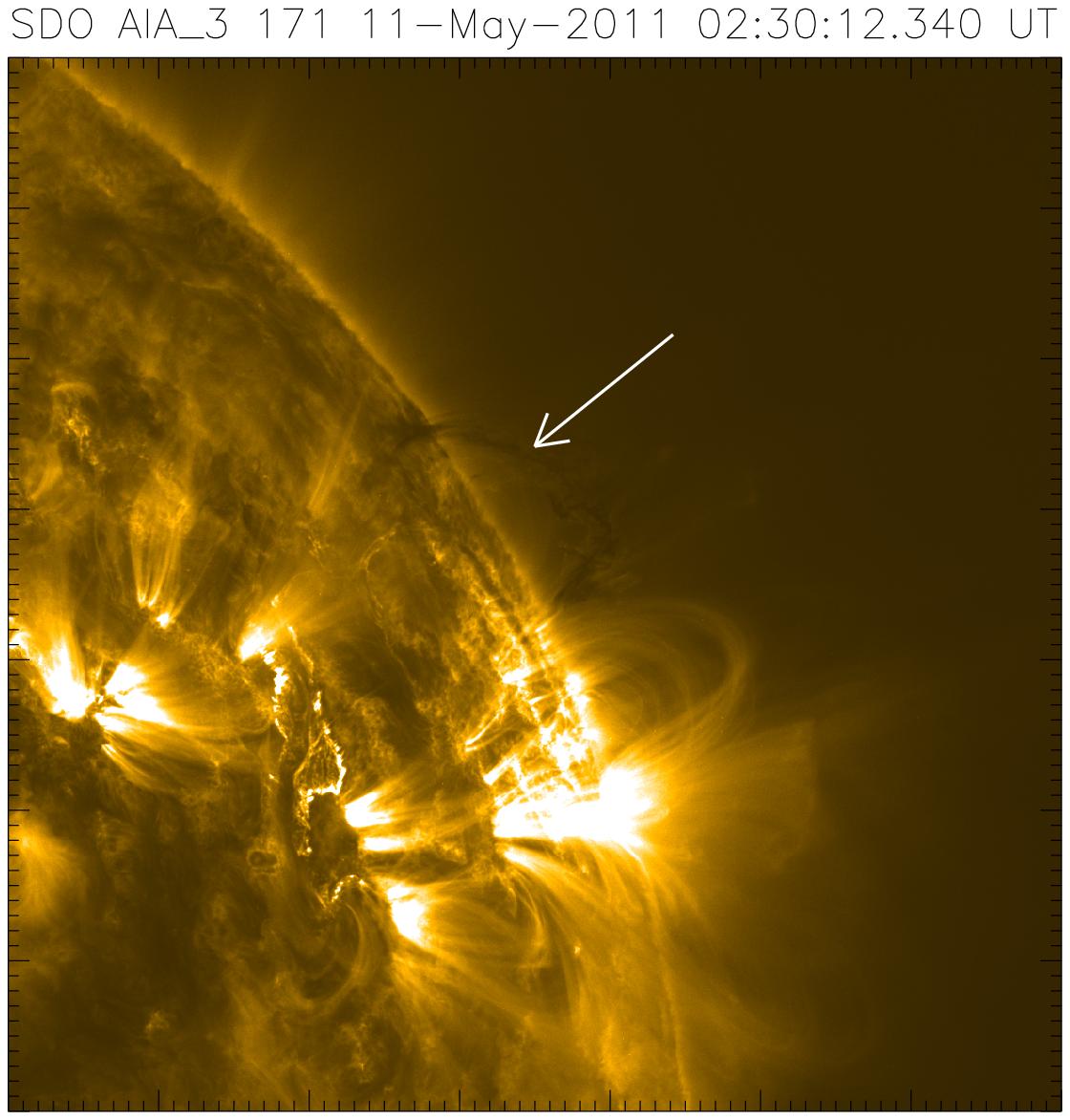}
}
\vspace{2mm}
\hbox{
\includegraphics[width=5.8cm]{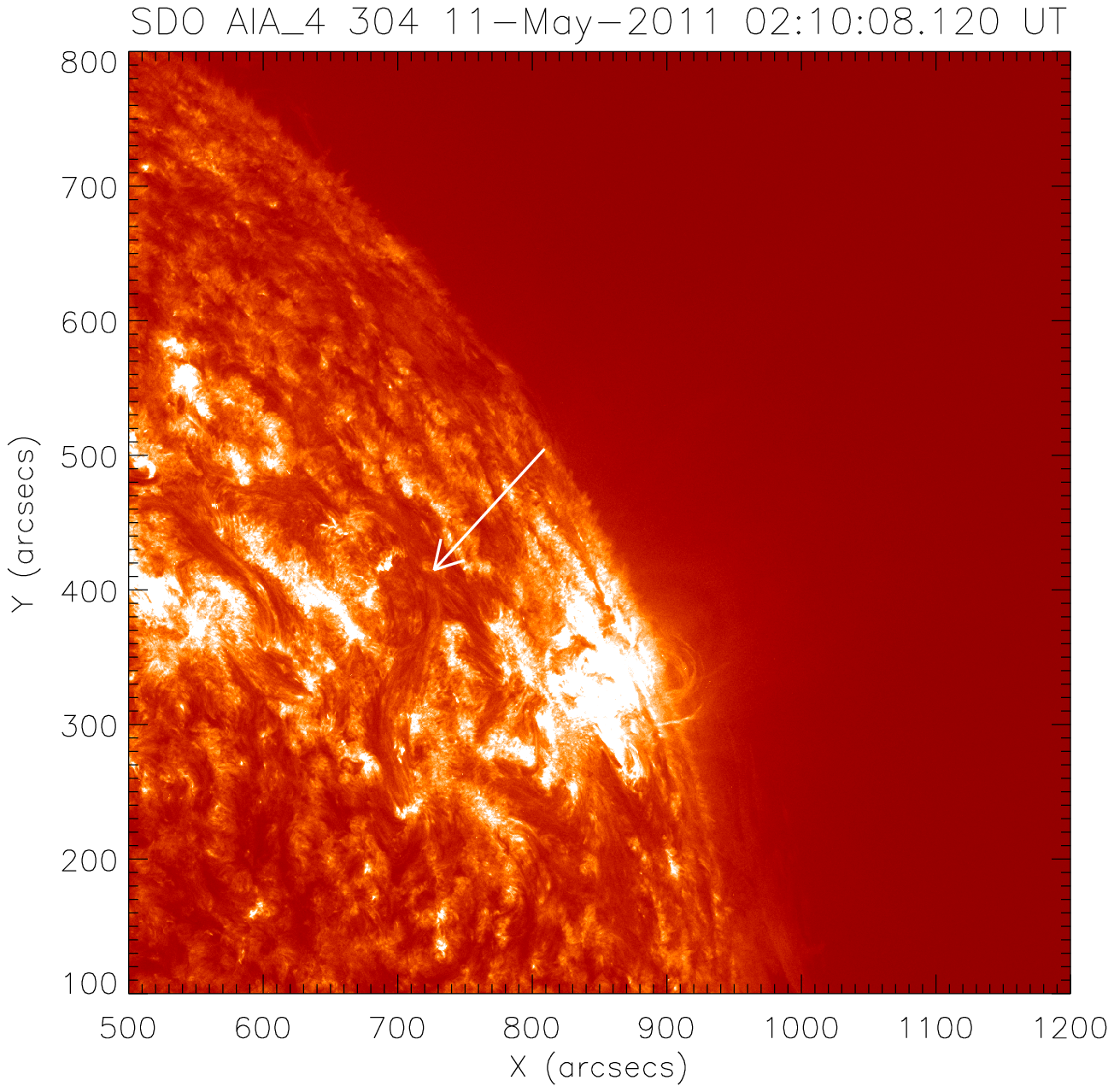}
\includegraphics[width=5.3cm]{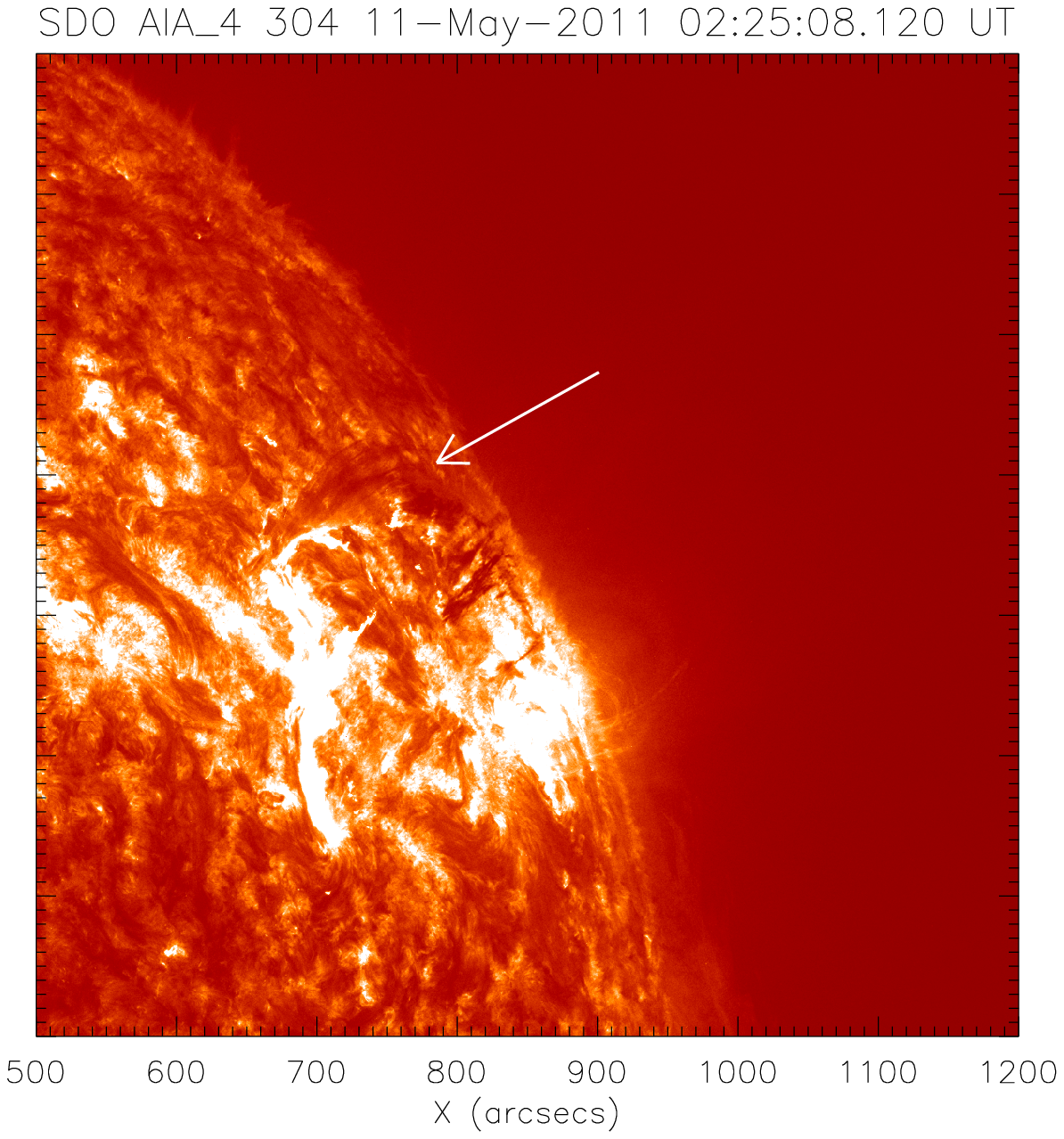}
\includegraphics[width=5.3cm]{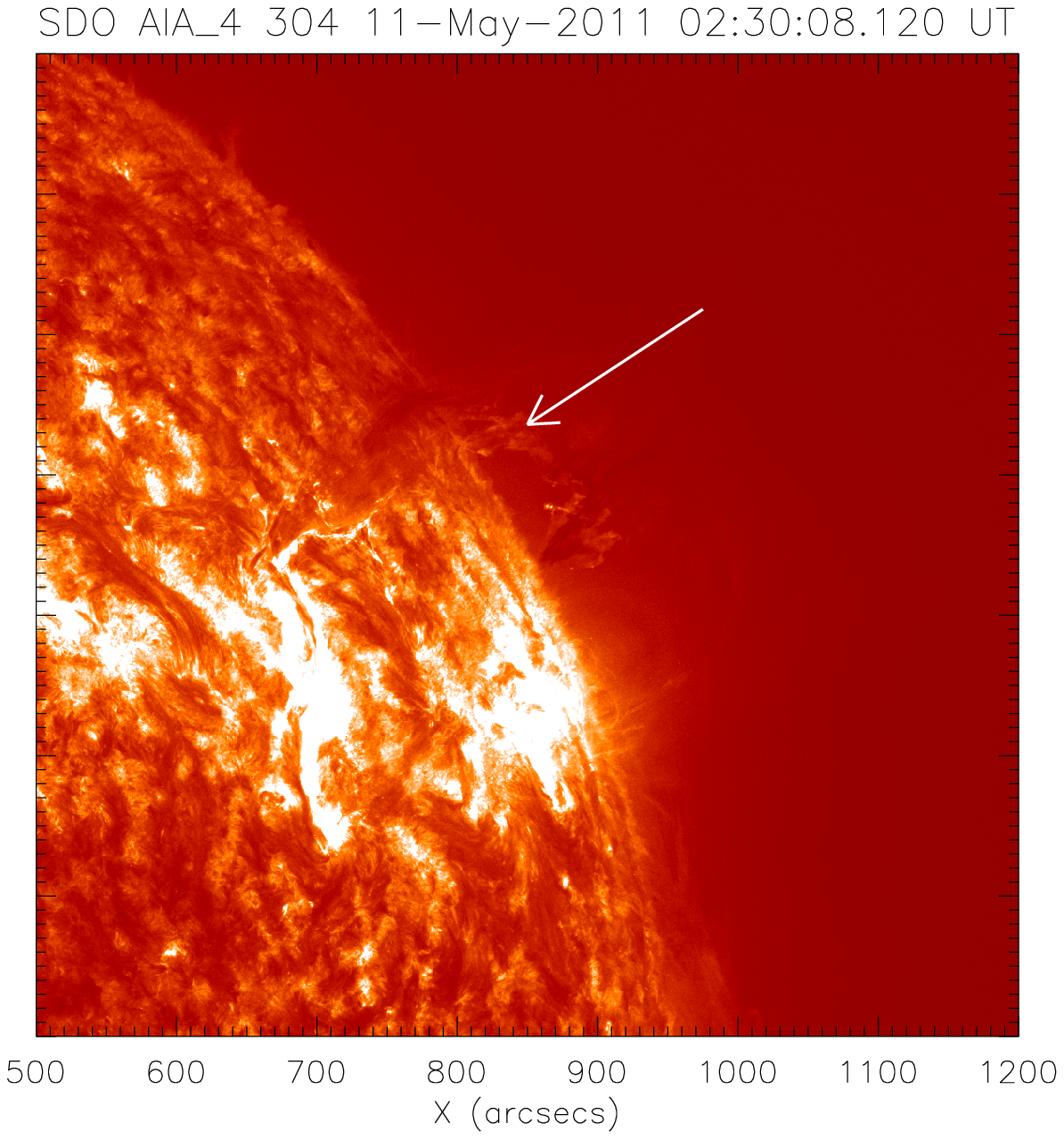}
}
  \end{minipage}
  \caption{Evolution of the filament eruption followed by the flare observed
by {\em SDO}/AIA in 171 \AA\ and 304 \AA, respectively.}
\label{sdo}
\end{figure}

\begin{figure}
\centering
\hbox
{
\hspace{-5.5cm}
\includegraphics[width=0.9\textwidth,viewport=  0 190 507 800, clip=]{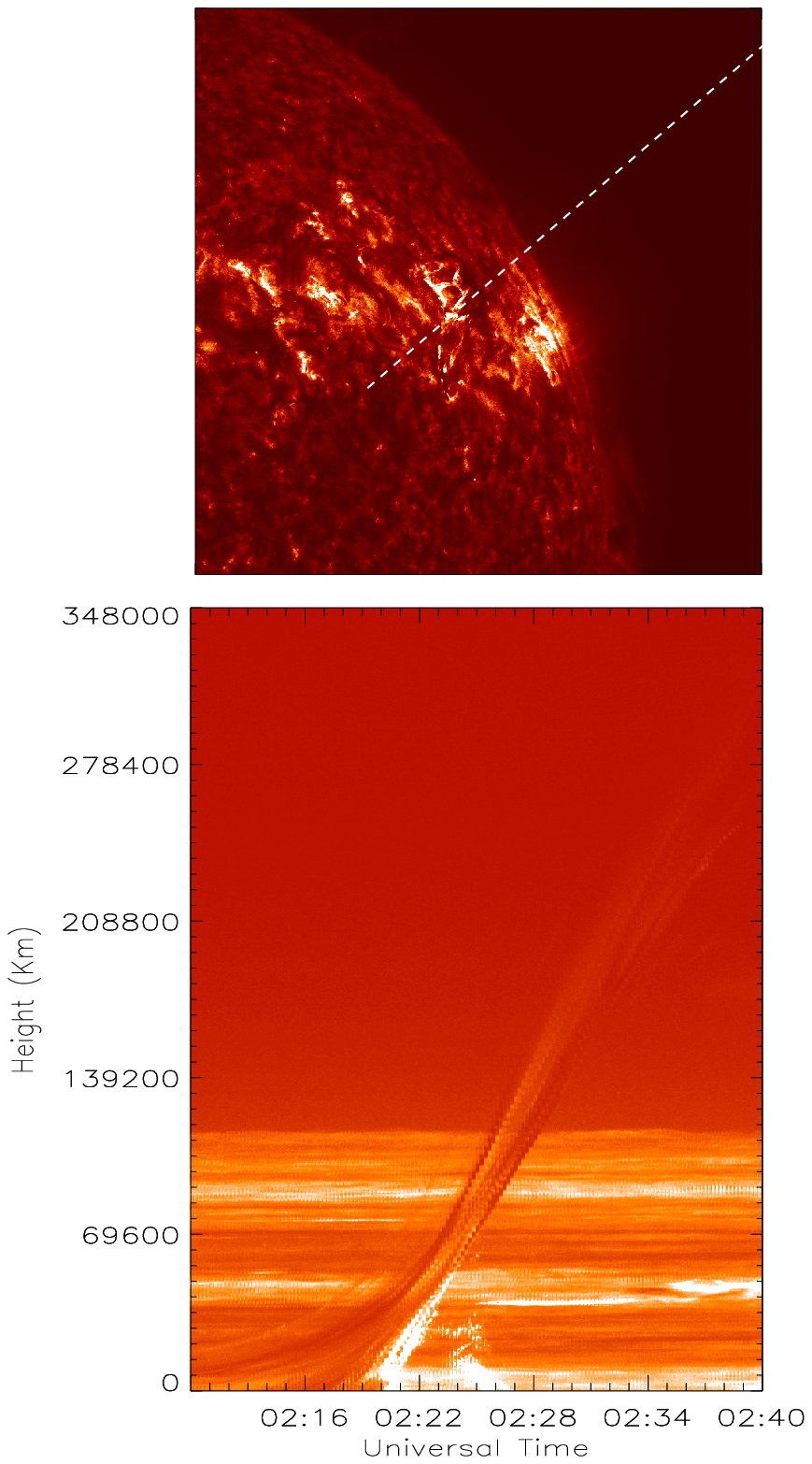}
\hspace{-4.0cm}
\includegraphics[width=0.75\textwidth,viewport=  0 -30 507 500, clip=]{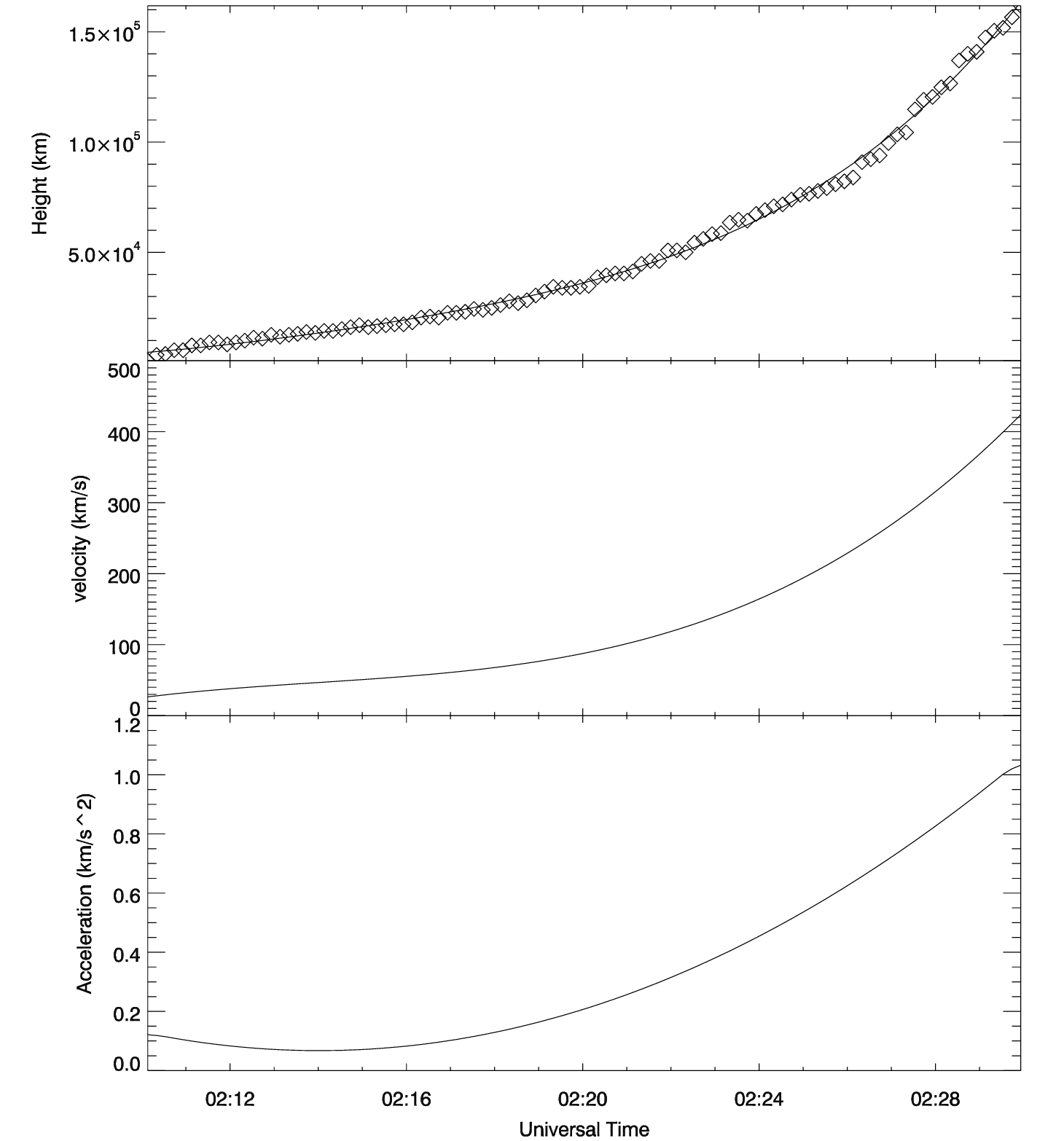}
}
\caption{{\it Top left}: the {\em SDO} 304 \AA\ image, where the slice is used
to plot the time-distance diagram for the erupting filament; {\it Bottom left}:
Time-slice diagram of the filament eruption; {\it Right top}: time evolution of
the filament height in observations ({\it diamonds}), which is fitted by a
fourth order polynomial function ({\it line}); {\it Right middle}: time
evolution of the filament velocity; {\it Right bottom}: time evolution of the
filament acceleration.}
\label{ht}
\end{figure}

\section{Filament Eruption and the Associated Phenomena}
\label{filament}

On 2011 May 11 the filament under study is located between the active regions
NOAA 11207 and NOAA 11205 at N20W60 on the solar disk. It has a length of
$\sim$150 Mm. To its north, there is another short filament, which shares the
same magnetic neutral line. During eruption, only the longer filament
erupts. This filament starts to rise at $\sim$02:10 UT on 2011 May 11.
The eruption of filament is followed by a weak flare. According to
\textit{Geostationary Operational Environmental Satellite (GOES)} observations, the flare is classified as B9.0-class. The soft
X-ray enhancement starts at around 02:20 UT, peaks at 02:40 UT and disappears
after 03:20 UT. Looking at the spatial evolution of flare in Figure
\ref{sdo}, the flare shows two quasi-parallel ribbons. As the filament moves
up, the two ribbons start to separate from each other as expected from the
standard CSHKP model. The ribbons are located on the opposite sides of the
magnetic neutral line. The filament eruption is associated with a CME.
According to the LASCO CME catalog, the CME appears in the LASCO field-of-view
around 02:48 UT . The CME is a partial halo event with an angular width of
225$^{\circ}$. The speed and the acceleration of the CME are 740 km s$^{-1}$
and 3.3 m s$^{-2}$, respectively.

In order to see the kinematics of filament eruption, we create a time-slice
diagram using the {\em SDO}/AIA 304 \AA\ data. The location of the slice is shown
in the top-left panel of Figure \ref{ht}, and the corresponding time-slice
plot is displayed in the bottom-left panel, according to which we plot the
time evolution of the filament height in the top-right panel of Figure
\ref{ht}. The diamond symbols correspond to the observations, which are nicely
fitted by a fourth order polynomial function. The fitted line is overplotted
on the observed data. From the height-time plot, we compute the velocity and
the acceleration of the erupting filament. The derived values are plotted in
the middle and bottom panels in the right column of Figure \ref{ht}. It is found
that velocity varies in the range of 30--400 km s$^{-1}$ and the estimated
acceleration varies from 0.1 to 1 km s$^{-2}$.

In many reported cases, filament eruptions often exhibit distinct two phases, i.e.,
slow and fast rise phases \citep{Chifor06, Schrijver08, Koleva12, Joshi13}. 
Interestingly, in our case, the velocity of the erupting filament changes
continually, and we cannot divide the evolution into two phases. Such
type of eruptions were proposed in the case of the kink instability
\citep{Torok05, Ji03, cheng12}. In this case, the eruption can occur without
the need of the slow rise phase with a nearly constant velocity. Therefore, our
eruption event may be initiated by the kink instability on the first instance.
However, we did not observe the clear number of twist which meets the
Kruskal-Shafranov condition for the kink instability \citep{Sri10}.

\begin{figure*}
\centering
\hspace*{-1cm}
\includegraphics[width=1.2\textwidth,clip=]{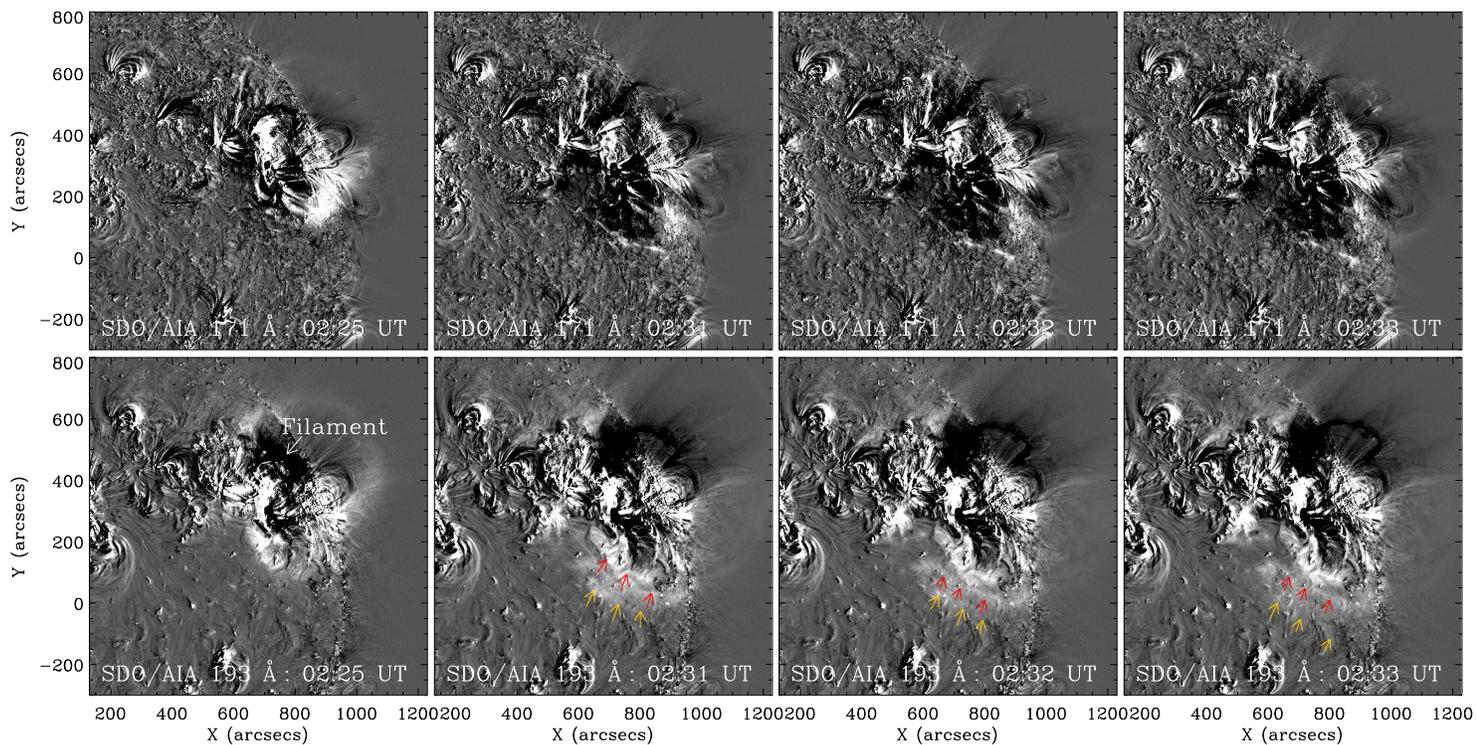}
\caption{Base difference images of the event observed in {\em SDO}/AIA 171 \AA\ 
and 193 \AA. The base image is taken at 02:00 UT. The yellow and red arrows 
indicate the fast-mode MHD wave and a slowly-moving EIT wave, respectively. 
The white arrow indicates the erupting filament.}
\label{evolution}
\end{figure*}

\section{Two Types of EUV waves and Stationary EUV Wave Fronts}
\label{euv}

The filament eruption on 2011 May 11 is associated with EUV waves. The first 
appearance of the EUV waves is around 02:00 UT. The wave is seen to propagate 
mostly in the south-west direction. We display the AIA 171 and 193 
\AA\ base difference images to see the evolution of the EUV waves in the two 
rows of Figure \ref{evolution}, respectively. To make the base difference 
images, a pre-event image at 02:00 UT is subtracted from each observed image. 
Propagating EUV waves are clearly seen in Figure \ref{evolution}, including a 
fast-moving EUV wave marked by yellow arrows and another slowly-moving EUV wave
indicated by red arrows. The slowly-moving EUV wave is followed by coronal 
dimmings in both wavelengths. Since the coronal dimmings can be 
observed in different wavelengths, they are mainly due to the depletion of plasma
density. Several authors quantitatively calculated the dimmings 
due to the plasma density depletion using {\em Yohkoh}/SXT \citep{Sterling97}, {\em SOHO}/EIT \citep{Zhukov04} 
and {\em STEREO} data \citep{Aschwanden09}. 

\begin{figure*}
\begin{minipage}[b]{0.5\linewidth}
\centering
$\color{black} \put(0,-50){\Large \bf (a)} $
$\color{black} \put(230,-50){\Large \bf (b)} $
\includegraphics[width=0.99\textwidth,clip=]{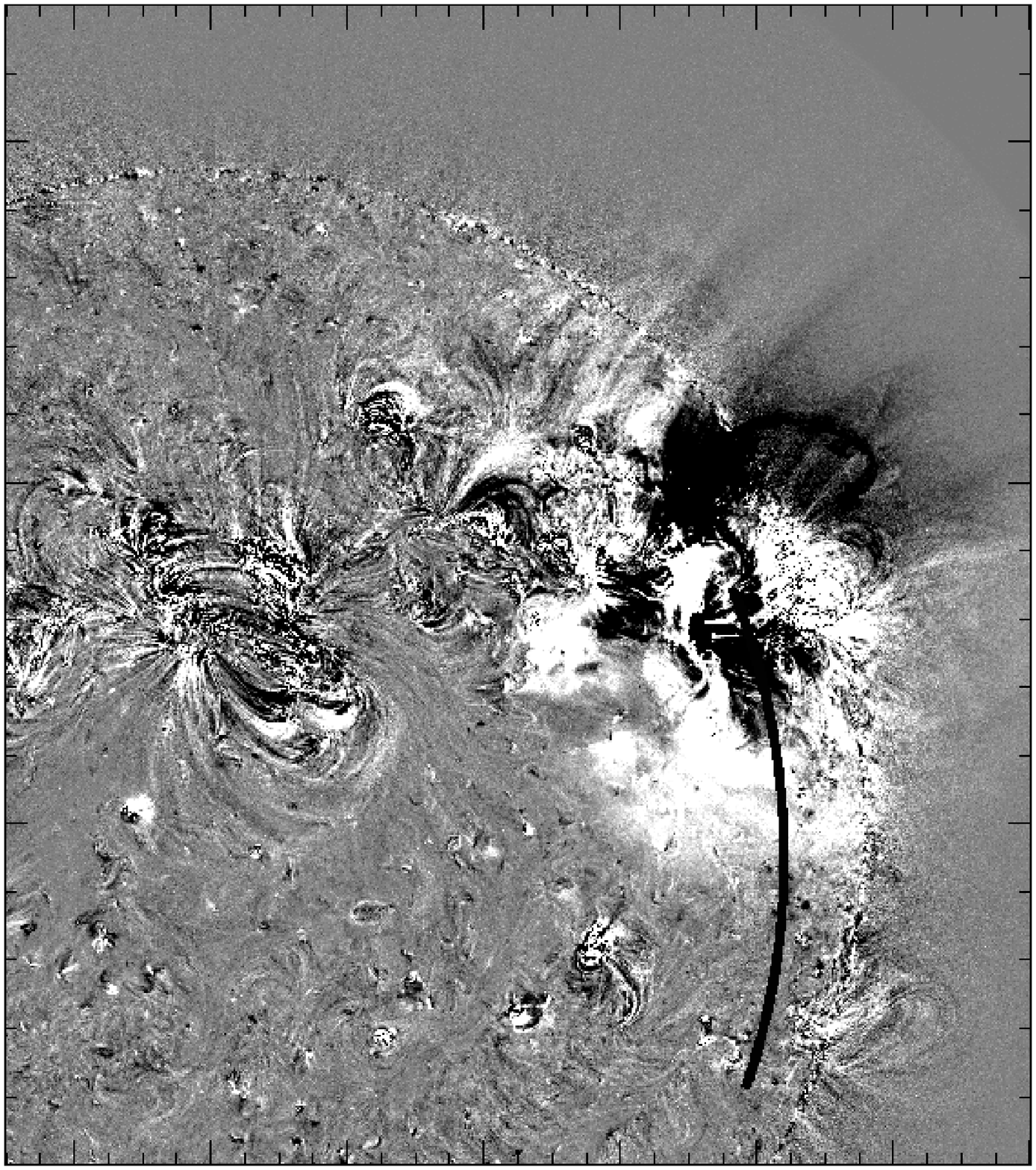}
\end{minipage}
\hfill
\begin{minipage}[b]{0.5\linewidth}
\centering
\hspace*{-1cm}
\vspace*{-0.7cm}
\includegraphics[width=1.18\textwidth,clip=]{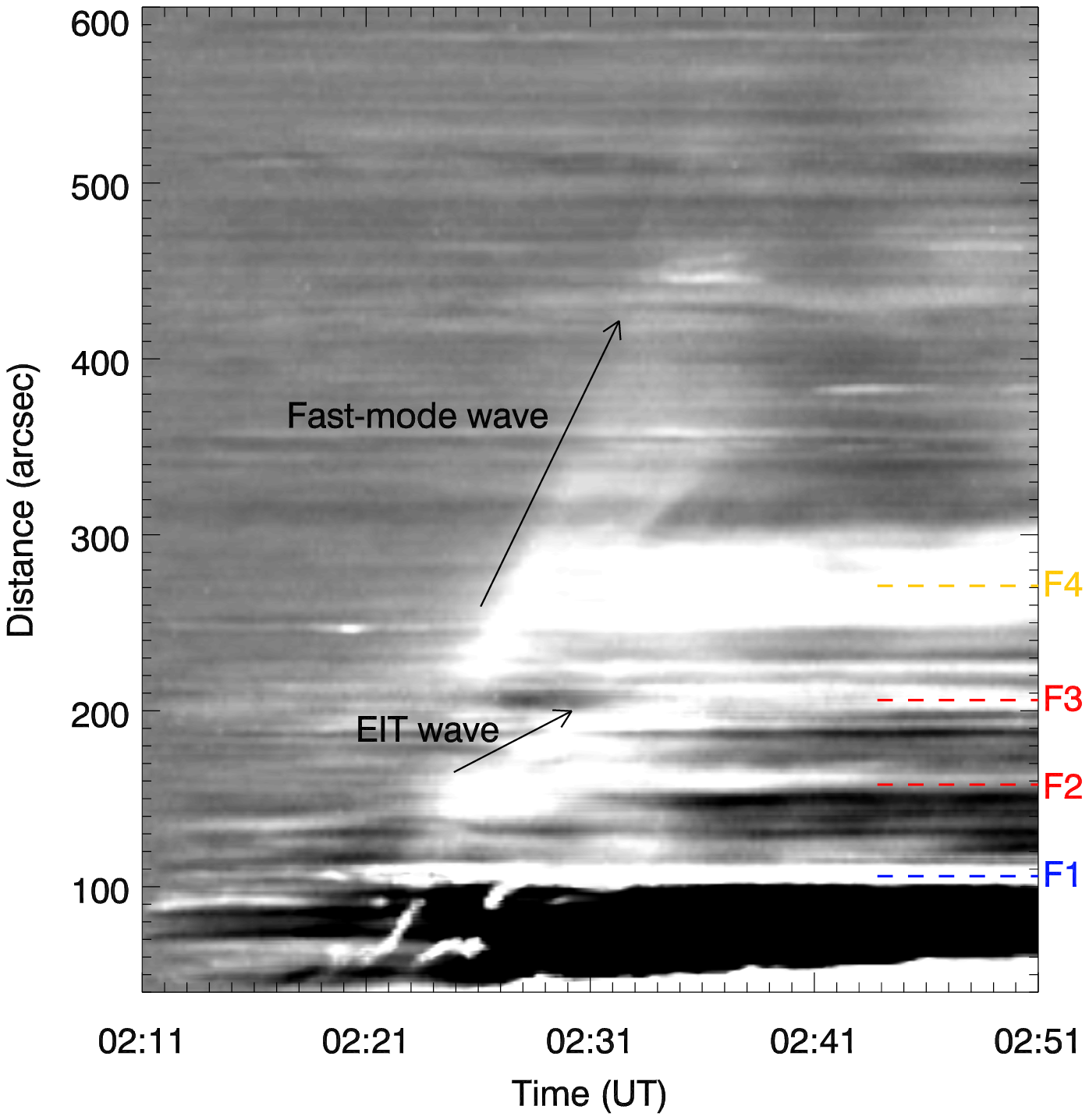}
\end{minipage}
\caption{(a): The {\em SDO}/193 \AA\ (02:11:07 UT) difference image
showing a slice ({\it black line}) to be used in the time-distance diagram. (b):
Time-distance diagram showing two types of EUV waves and several stationary
fronts, $F_1$, $F_2$, $F_3$, and $F_4$. The fast-mode MHD wave and the
slowly-moving EIT wave are marked by the arrows.}
\label{slice}
\end{figure*}

To see the kinematics of the EUV waves clearly, we create a time-slice image
in AIA 193 \AA. As shown in the left panel of Figure \ref{slice}, the slice is
a great circle starting from the flare region. The right panel of Figure
\ref{slice} displays the time evolution of the 193 \AA\ intensity distribution
along the slice. Inspecting the time-slice diagram, we find that there are two
types of waves, one is a fast-moving wave and another is a slowly-moving wave.
We claim the fast-moving wave as the fast-mode MHD wave and the slowly-moving
wave as the EIT wave, as marked by the arrows in the right panel of Figure
\ref{slice}. The speed of the fast-mode wave is $\sim$500 km s$^{-1}$, which is
several times greater than the coronal sound wave. The observed slower wave is
a typical EIT wave. The initial speed of the EIT wave is $\sim$120 km s$^{-1}$,
which is even smaller than the coronal sound speed. Note that with the AIA 193
\AA\ formation temperature, the coronal sound speed is 186 km s$^{-1}$. It is
also seen that as time progresses, the foremost front of the fast-mode wave
keeps a constant speed, whereas the speed of the EIT wave decreases, and around
02:33 UT it stops. Since the EIT wave bifurcates into two fronts, they form
two stationary fronts, $F_2$ and $F_3$ at distances of 160\arcsec\ and
200\arcsec, respectively. Since {\em SDO}/AIA has a 12 s cadence, we can follow
the propagation of any EUV waves. It is seen that the stationary fronts at the
distances of 160\arcsec\ and 200\arcsec\ in Figure \ref{slice}(b) indeed
result from the gradual deceleration of the slowly-moving EIT waves.

\begin{figure*}
\centering
\includegraphics[width=0.9\textwidth,clip=]{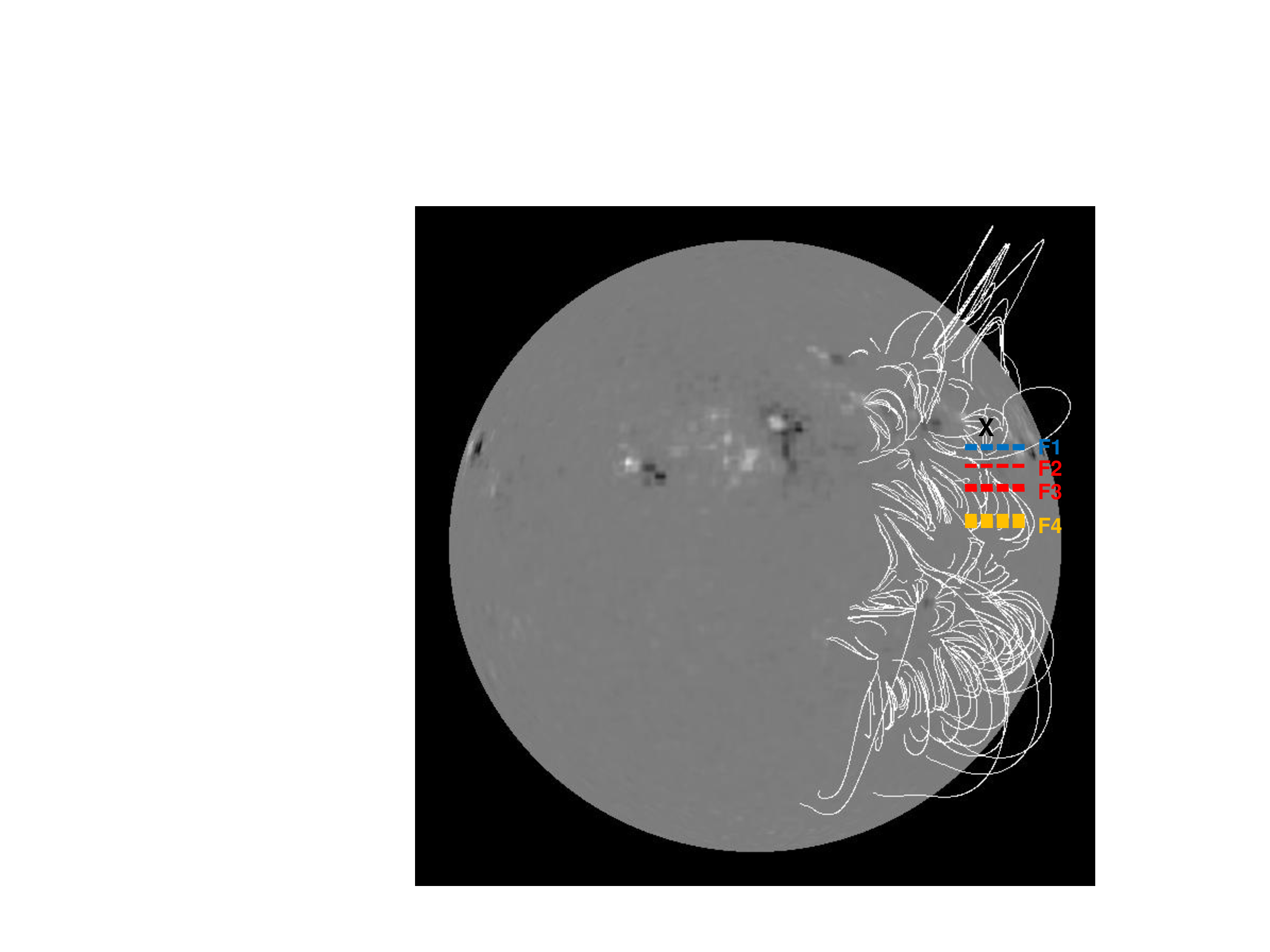}
\caption{Extrapolated coronal magnetic field using the Potential Field
Source-Surface (PFSS) model. The photospheric magnetic field is taken at 00:04
UT observed by HMI on board {\em SDO}. The locations of several stationary fronts shown in 
figure 4 are overploted. The black cross indicates the location of the active region.}
\label{extra}
\end{figure*}

More interestingly, we notice another two stationary fronts in Figure
\ref{slice}(b), which are not related to the EIT waves. The first one, $F_1$,
is located at a distance of 110\arcsec, and the second one, $F_4$, is at a
distance of 280\arcsec\ in Figure \ref{slice}(b). The first one, which is very
close to the flare site, is the border of a core dimming region. In order to
understand the formation of these stationary EUV fronts, we plot the
extrapolated coronal magnetic field in Figure \ref{extra}, where the
extrapolation is based on the potential field source surface (PFSS) model.
After checking the extrapolated magnetic field, we find that Front
$F_2$ (marked by the red line) is nearly cospatial with magnetic separatrice
or quasi-magnetic separatrix layers (QSLs) where magnetic field lines diverge 
rapidly. Note that a magnetic separatrix is a special case of magnetic QSL, 
where the neighboring magnetic fields belong to different magnetic systems. 
Front $F_4$ (marked by the yellow line) is not cospatial but very close to
another QSL. Beside, Front $F_1$ is located inside the magnetic system of the
source region, and Front $F_3$ is shifted slightly from the QSL that is nearly
cospatial with Front $F_2$.

\begin{figure*}[h]
\centering
\includegraphics[width=0.8\textwidth,clip=]{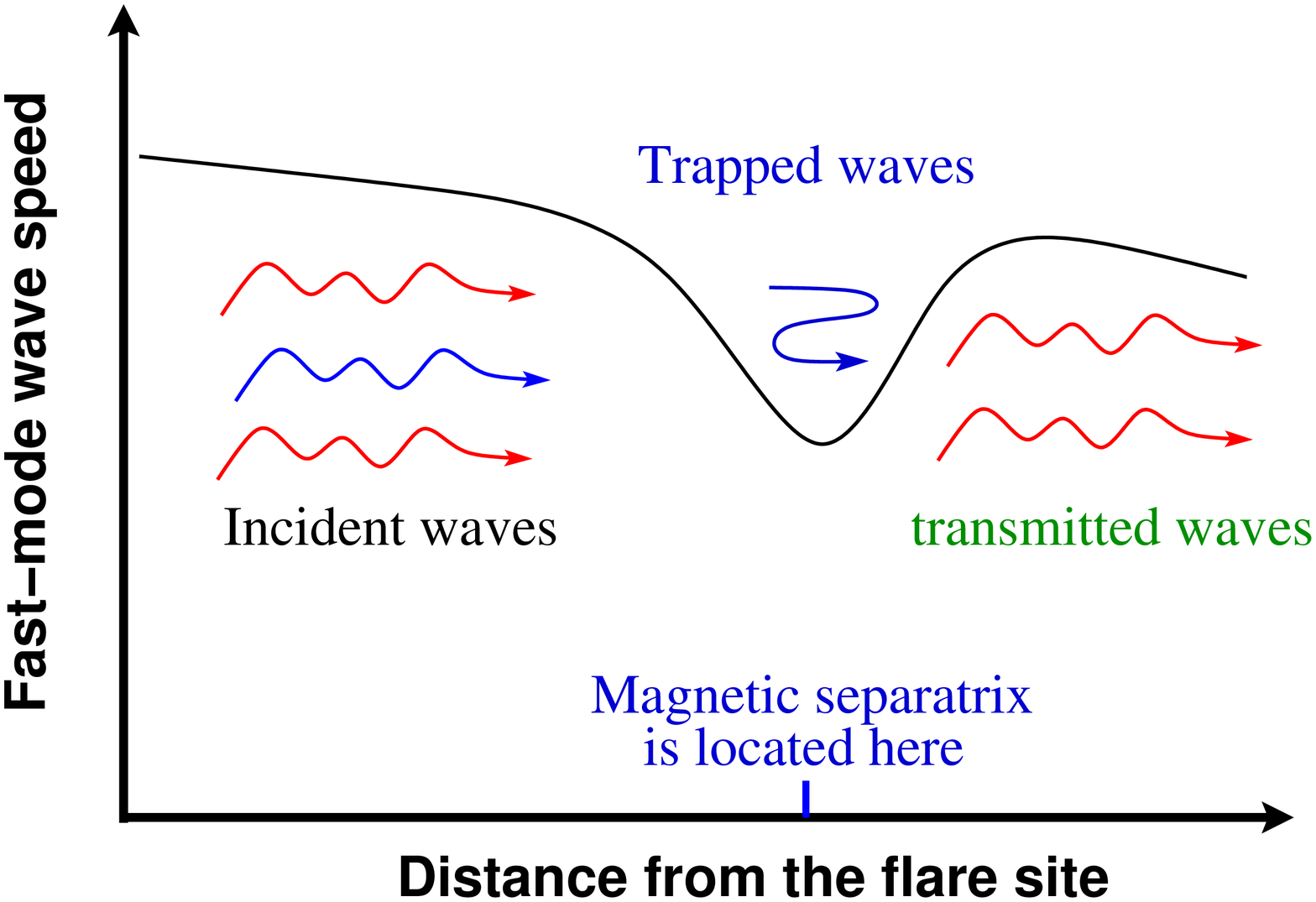}
\caption{Illustration of physical process for the interpretation of the stationary front $F_4$.}
\label{cartoon}
\end{figure*}

\section{Discussion}
\label{discussion}

When EIT waves were discovered, they were initially considered as fast-mode
MHD waves \citep{Thompson98,Wang00, Wu01}, i.e., they are long-awaited
coronal counterparts of chromospheric Moreton waves. Moreton waves were
discovered by \citet{Moreton60} and \cite{Moreton60b} as a dark front followed
by a bright front in the H$\alpha$ red wing or a bright front followed by a
dark front in the H$\alpha$ blue wing. They have a typical velocity of the order
of 1000 km s$^{-1}$ \citep{Smith71}. Despite some apparent evidence that seems
to support the fast-mode wave nature of EIT waves \citep[e.g.,][]{Olmedo12,
Gopalswamy09, Ballai05}, a serious problem with the fast-mode wave model is
that the EIT wave speed is typically $\sim$3 times slower than Moreton waves
\citep{Klassen00}, and in some cases the EIT wave speed is only $\sim$80 km s
$^{-1}$ \citep{Klassen00} or even $\sim$10 km s$^{-1}$ \citep{zhuk09}. 
\citet{Nitta13} claimed that the large-scale coronal propagating fronts 
have a mean wave speed of 644 km s$^{-1}$, which is comparable to that of
Moreton waves. However, they always selected the fastest front in each of their 
time-slice diagrams. Therefore, in our view, most events in their paper are the 
coronal counterpart of Moreton waves, rather than the original EIT waves found 
by \citet{Thompson98}. In our study, whenever we say that EIT wave is 
generally three times slower than the fast-mode wave in the corona, we mean the
slower one in the two-wave paradigm.

The lack of correspondence between the speeds of Moreton and EIT waves was
also suggested by \citet{Warmuth01, Warmuth04a, Warmuth04b}. In order to explain the velocity difference, they proposed that 
fast-mode wave decelerates from typical Moreton wave speeds to typical EIT wave
speeds. Whereas this idea may be able to explain the deceleration of the real
fast-mode wave, whose speed is higher near the source active region than in the
quiet region, we definitely need another model to explain those EIT waves whose
speeds are below the sound speed. In order to explain the low speeds of many
typical EIT waves, \citet{Chen02} and \citet{Chen05} proposed that two types 
of EUV waves are formed in association with a filament eruption. The 
fast-moving wave is a piston-driven shock wave, which corresponds to the 
coronal counterparts of the chromospheric Moreton waves, and the slowly-moving 
wave is an apparent motion, which is formed due to the successive stretching of 
the closed magnetic field lines overlying the erupting flux rope.

The co-existence of two types of EUV waves was initially verified by
\citet{Harra03}, and later conclusively confirmed by {\em SDO}/AIA observations
\citep{Chen11, Asai12, Kumar13, White13}. It can also be identified in many
events statistically analyzed by \citet{Nitta13}. As predicted by the magnetic
field-line stretching model \citep{Chen02}, the fast-moving EUV wave is
about 3 times faster than the slowly-moving EUV wave. For example, the ratio
is 2.5 \citep{Harra03}, 2.9 \citep{Chen11}, 3.4 \citep{Kumar13}, and 1.8
\citep{White13}. In this paper, we also found two EUV waves, with the velocity
ratio being 4.2. According to the magnetic field-line stretching model, this
relatively large ratio implies that the closed field lines overling the filament
are relatively more stretched in the radial direction. Besides, it is
seen that expanding dimmings immediately follow the slower EIT wave, as 
illustrated by Figure \ref{evolution}. Such a feature is again consistent with the
magnetic fieldline stretching model, which interprets that the EIT waves and 
expanding dimmings are both due to the field line stretching.

Another feature of EIT waves that led to the doubt on the fast-mode wave model
is the stationary fronts. \citet{Delannee99} first reported that an EIT brightening
remains at the same location for tens of minutes. They called such brightenings
as ``stationary brightenings". Later on, such stationary brightenings were
confirmed in several observational studies \citep{Delannee00,Delannee07,
Attrill07, Chandra09}. Such a stationary front located at a magnetic separatrix
or a QSL in more general cases was reproduced in numerical simulations, and can
be well explained by the magnetic field-line stretching model \citep{Chen05,
chen06a}. Even though, there has still been doubt about the validity of the
stationary fronts due to the low cadence of the EIT telescope. With the high
cadence observations of the 2011 May 11 event by {\em SDO}/AIA, we confirm that
the slowly-propagating EIT wave finally stops at a magnetic QSL. One peculiar
feature in this event is that the EIT wave bifurcates into two stationary
fronts, $F_2$ and $F_3$ in the time-slice diagram (Figure \ref{slice}), 
and only the first front, $F_2$, is cospatial with a QSL, with the
other one being slightly shifted away. These
detailed structures cannot be detected with the telescopes
before {\it SDO} was launched. One possibility of the bifurcation is that the 
outer front is the traditional EIT wave front, whereas the inner front is 
simply an expanding coronal loop, as proposed by \citet{cheng12}. Another 
possibility, which we favor, is that the two fronts are due to the projection of different 
layers of one EIT wave front since the EIT wave front has a domelike structure 
in 3-dimensions \citep{vero10}. In addition to
the bifurcation of the slower EUV wave into fronts $F_2$ and $F_3$, even inside
front $F_3$, a multitude of strands are identifiable. One might wonder whether
the fine structures inside front $F_3$ can be explained by slow-mode shocks,
which was proposed by \citet{wang09, wang15}. With the current observations, we
could not tell. As for the soliton model \citep{will07}, we are still not sure
whether a slow-mode soliton wave can propagate across magnetic field lines and
stop at magnetic separatrix. More strikingly,
 we find two more stationary fronts $F_1$ and
$F_4$, where $F_1$ is close to the flare site, and $F_4$ is formed when the 
fast-mode wave interacts with another magnetic QSL.

It seems from Figure \ref{slice} that the stationary front $F_1$ emanates at
$\sim$02:19 UT, which is slightly earlier than the onset of the solar flare around 02:20 UT.
Therefore, it would be more related to the initiation of the filament eruption. 
It is noticed that this stationary front is located at the boundary of the core dimmings. 
Since the core dimmings are generally believed to be due to the evacuation of plasma
associated with the erupting flux rope \citep{Sterling97, Jiang03}, this stationary front
might be formed at the interface between the flux rope (near the footpoint) and
the envelope magnetic field (which is more like potential). In this sense,
the current shell model proposed by \citet{Delannee08} may provide a sound
explanation for front $F_1$. 

As for another stationary front $F_4$, it seems that it is formed when the 
fast-mode MHD wave passes through the magnetic QSL at a distance of 280\arcsec\
 away from the flare site. This feature has never been reported, and generally 
it is thought that a fast-mode wave may pass through a magnetic QSL freely, 
leaving no significant traces behind since a magnetic QSL is a topological 
characteristic, and the magnetic field strength may change smoothly across the 
QSL.  

However, from a theoretical point of view, when a wave propagates in a 
non-uniform medium, wave reflection would be produced when the intrinsic wave 
speed of the medium changes rapidly. In particular, when the wave speed in a 
layer is much lower than that of other regions outside, a wave passing through 
would be decomposed into a transmitting component and a trapped component that 
bounces back and forth inside this layer, just like the Fabry-P\'erot 
interferometer. Inspired by the observational result presented in this paper, 
we (2016, in preparation) call such a layer as ``magnetic 
valley", and are planning to study how such a magnetic valley responds to an 
incident wave by numerical simulations. So far a similar phenomenon was 
numerically investigated by \citet{Yuan15}. \citet{Murawski01} and
\citet{Yuan15} did a one-dimensional simulation of propagation of fast 
magnetoacoustic pulses in a randomly structured plasma and found that the 
magnetoacoustic pulses were trapped by the randomly structured plasma. 
Such a ``magnetic valley" exists when the QSL is a magnetic separatrix, and the
magnetic fields on the two sides of the separatrix belong to two different
magnetic systems. The magnetic field around the separatrix might be strongly
divergent. In this case, after a fast-mode MHD wave enters this magnetic valley,
only a part of the wave can be refracted from the low-Alfv\'en speed region
out to the high-Alfv\'en speed region, with the remaining part of the wave 
being trapped in the magnetic valley, bouncing back and forth between the two 
interfaces. We illustrate this physical process in Figure \ref{cartoon}. 
Definitely this observational feature of EUV waves merits further numerical 
simulations. Unfortunately we cannot identify the bouncing waves at the
stationary front. The possible reason is that in 2 or 3 dimensions, the 
magnetic valley have different widths at different heights, contrary to the 
one-dimensional case. Therefore, trapped waves with different periods are mixed
together in the projected plane, making each wave not identifiable. It is also
noted that, as seen in Figure \ref{extra}, the stationary front $F_4$ is not
exactly cospatial with the QSL. Such a shift might be due to the limitation
of the PFSS model, or such a stationary front is formed with a mechanism 
different from our conjecture mentioned above.

\citet{Kwon13} also reported stationary EUV fronts after the passage of a
fast-mode shock wave. However, the two fronts in their paper are actually
separating in opposite directions with a small velocity. Since the two EUV
fronts are located on the two footpoints of a helmet streamer, the
brightenings are probably produced by the magnetic reconnection of the current
sheet above the helmet streamer triggered by the passing shock wave (B.
Vr{\v s}nak, private communication), which are like flare ribbons and
different from ours.

Alternatively, the formation of the stationary front $F_4$ might be interpreted
as stoppage of expansion of structures inside the CME as suggested by
\cite{cheng12}. \cite{cheng12} presented the study of formation and separation
of two EUV waves from the expansion of a CME. They also reported that the CME
and the faster EUV wave propagate with different kinematics after they decouple.

\section{Conclusions}
\label{sum}

In this study, we presented the observations of two propagating EUV waves,
i.e., a fast-mode MHD wave and a slowly-moving EIT wave associated with a
filament eruption and a CME, as found in many other CME events via {\it
SDO}/AIA. In association with the two propagating waves, we observed four
stationary fronts i.e., $F_1$, $F_2$, $F_3$, and $F_4$, as indicated by Figure
\ref{slice}.  The stationary wave fronts $F_2$ and $F_3$ are the results of the
gradual deceleration of the slowly-moving EIT wave, which finally stops near
the location of a QSL. The formation of Front $F_2$ can be 
explained by the magnetic field-line stretching model proposed by 
\citet{Chen02, Chen05}. Front $F_3$ is bifurcated from front $F_2$, so it is
shifted slightly away from the QSL. This might be due to the projection 
effects, i.e., Front $F_3$ is from a higher layer of the same domelike EIT
wave front as Front $F_2$.

Front F1 is proposed to be related to the initiation of the filament 
eruption and is located at the edge of the core dimmings. This may correspond 
to the edge of the erupting flux rope at the footpoint. It might be explained 
by the current shell model proposed by \citet{Delannee08}. Stationary front 
$F_4$ is observed for the first time. We tentatively explain it to be
formed when the fast-mode MHD wave interacts with a magnetic QSL. During the
interaction, a fraction of the wave passes through, with the rest being
trapped locally. Other possibilities are not excluded though.

\acknowledgments
We are thankful to the referee for his/her detailed comments and suggestions, which improved the manuscript significantly. The authors thank the open data policy of the SDO team. RC, AF is
supported by the ISRO/RESPOND project no. ISRO/RES/2/379/12-13, and PFC is
supported by the Chinese foundations (NSFC grants nos. 11533005 and 11025314).
\bibliographystyle{apj}
\bibliography{references}
\clearpage
\end{document}